\documentclass[twocolumn,floatfix,superscriptaddress,pra,aps,10pt,nofootinbib,longbibliography]{revtex4-2}
\usepackage{times}
\usepackage{amssymb}
\usepackage{color}
\usepackage{amsmath}
\usepackage{amsbsy}
\usepackage{amsthm}
\usepackage{graphicx}
\usepackage{bm}
\usepackage{epsfig}
\usepackage{xfrac}
\usepackage{xcolor}
\usepackage{enumerate}
\usepackage{multirow}
\definecolor{myurlcolor}{rgb}{0,0,0.7}
\definecolor{myrefcolor}{rgb}{0.8,0,0}
\usepackage[unicode=true,pdfusetitle, bookmarks=true,bookmarksnumbered=false,bookmarksopen=false, breaklinks=false,pdfborder={0 0 0},backref=false,colorlinks=true, linkcolor=myrefcolor,citecolor=myurlcolor,urlcolor=myurlcolor]{hyperref}

\newcommand{\ket}[1]{\left| {#1} \right\rangle}
\newcommand{\bra}[1]{\left\langle {#1}\right|}

\newcommand{\braket}[2]{\langle #1|#2\rangle}

\renewcommand{\t}[1]{\textrm{#1}}
\newcommand{\tr}[0]{\mathrm{Tr}}

\usepackage{braket}
\newcommand{\thmref}[1]{\hyperref[#1]{Theorem~\ref{#1}}}
\newcommand{\lemmaref}[1]{\hyperref[#1]{Lemma~\ref{#1}}}
\newcommand{\figref}[1]{\hyperref[#1]{Fig.~\ref{#1}}}
\newcommand{\figaref}[1]{\hyperref[#1]{Fig.~\ref{#1}a}}
\newcommand{\figbref}[1]{\hyperref[#1]{Fig.~\ref{#1}b}}
\newcommand{\figcref}[1]{\hyperref[#1]{Fig.~\ref{#1}c}}
\renewcommand{\eqref}[1]{\hyperref[#1]{(\ref{#1})}}
\newcommand{\eqsref}[2]{\hyperref[#1]{Eqs.~(\ref{#1})-(\ref{#2})}}
\newcommand{\appref}[1]{\hyperref[#1]{Appx.~\ref{#1}}}

\newcommand{\var}{\alpha}

\def\labell#1{\label{#1}}
\def\>{\rangle}\def\<{\langle}
\def\togli#1{}

\begin{document}
\title{Quantum metrology of noisy spreading channels}
\author{Wojciech G{\'{o}}recki}
\affiliation{Faculty of Physics, University of Warsaw, Pasteura 5, 02-093 Warsaw, Poland}
\author{Alberto Riccardi}
\author{Lorenzo Maccone}
\affiliation{Dip.~Fisica and INFN
    Sez.~Pavia, University~of Pavia, via Bassi 6, I-27100 Pavia,
    Italy}
\begin{abstract}
  We provide the optimal measurement strategy for a class of noisy
  channels that reduce to the identity channel for a specific value of
  a parameter (spreading channels). We provide an example that is
  physically relevant: the estimation of the absolute value of the
  displacement in the presence of phase randomizing noise.
  Surprisingly, this noise does not affect the effectiveness of the
  optimal measurement. We show that, for small displacement, a
  squeezed vacuum probe field is optimal among strategies with same
  average energy. A squeezer followed by photodetection is the optimal
  detection strategy that attains the quantum Fisher information,
  whereas the customarily used homodyne detection becomes useless in
  the limit of small displacements, due to the same effect
    that gives Rayleigh's curse in optical superresolution.  There is
  a quantum advantage: a squeezed or a Fock state with $N$ average
  photons allow to asymptotically estimate the parameter with a
  $\sqrt{N}$ better precision than classical states with same energy.
\end{abstract}

\maketitle
The goal of quantum metrology \cite{review,reviewrafal} is twofold:
(1) estimate the ultimate limits in the estimation of a parameter
$\var$ that is encoded into a physical probe by some transformation
or channel $\Lambda_\var$, and (2) find the optimal strategies that
attain this limit, namely the ones that achieve the quantum Fisher
information on the optimal probe.  In the noiseless case, where
$\Lambda_\var$ is a unitary transformation, the ultimate limits (the
Heisenberg bound) and the optimal estimation strategies are known, and
a quantum advantage exists either through entanglement \cite{qmetr} or
squeezing \cite{squeezemet}, typically a quadratic enhancement of
$\sqrt{N}$ in precision, where $N$ is the number of entangled probes
or is the average number of photons (or energy) employed in the
estimation.  In the noisy case \cite{rafalnoise,davidovich}, the
situation becomes very complicated and noise-dependent: there are many
transformations for which all quantum advantage is lost
\cite{rafaldephasing} and the optimal detection strategy is known only
for a handful of them \cite{reviewrafal}. In this paper we obtain a
local optimal detection strategy for a large class of noisy channels,
and we show that a physically relevant one of this class retains the
usual $\sqrt{N}$ quantum advantage. These are channels parametrized by non-negative parameter $\var\geq 0$ that morph into
the identity channel for $\var=0$. We call them ``spreading channels'', since the noise is
increased as the parameter increases. More rigorously, a spreading
channel $\Lambda_\var$ is defined as having the property
$\lim_{\var\to0}\Lambda_\var[\rho]=\rho$, with $\Lambda_\var$
differentiable in $\var=0$. Our strategies are optimal in the
proximity of $\var=0$.

The spreading channels may be seen as the ones obtained by the action
of a unitary $U_{\var,\varphi}=U^\dagger_\varphi e^{i\var G}U_\varphi$
with random, rapidly varying directions $\varphi$, distributed
according to a distribution $p(\varphi)$, which for specific cases was
discussed in \cite{Nichols2016,smerzi}. A related, but different,
problem refers to the ``nuisance parameters''\cite{masahito,nuisance},
where one works under the assumption that the uninteresting (nuisance)
parameter $\varphi$ has values {\em close} to some {\em known}
value. We drop this assumption here. Note that our analysis is also
different from the problem of quantum estimation in the absence of a
reference frame
\cite{Banaszek2004,Barlett2007,Fanizza2021squeezingenhanced}, which
can be seen as the action of a rotation $U_\varphi$ on the final state
(we, in contrast, consider a random rotation of the channel itself).
For the class of the channels discussed in this paper we show that the
averaging over the parameter $\varphi$ does not affect the efficiency
of extracting the information about the parameter $\var$ from the
output state.

A physically relevant example of discussed class channels is the
estimation of a small value of a displacement
$D(\var,\varphi)=e^{\var(e^{i\varphi} a^\dag-e^{-i\varphi}a)}$ of a
mode $a$ of the electromagnetic field in the presence of complete
randomization over $\varphi$ \cite{smerzi,gerardo}. This is relevant
for many estimation procedures, such as for axion dark matter searches
\cite{backes2021quantum,teufel2009nanomechanical,dassonneville2021dissipative,roni},
in communication channels with OOK modulation with dephasing, in
magnetic field estimation either at high temperature
\cite{Nichols2016} or in the presence of a trapped ion with unknown
phase \cite{smerzi}, in gravitational wave detection with resonant
cavities, e.g.~\cite{gemma}. Similar issues appear in many optical
imaging procedures \cite{mankey,mankey1,mankey2,aephr}. \togli{Indeed, the putative
  action of the axion field is to induce a (small) displacement on the
  electromagnetic field \cite{coreani,caterina}, but the phase
  $\varphi$ of the displacement is equal to the phase of the axion
  field, and hence it is quickly randomized on a time scale of the
  axion coherence time. [This is due to the fact that the putative
  axion field is a classical field that weakly couples coherently to
  the em field.] The action of the axion field is the channel
  $\Lambda_\var[\rho]=\int d\varphi D(\var,\varphi)\rho
  D^\dag(\var,\varphi)/2\pi$: a nonzero value of $\var$ is an
  indication of the presence of the axion field. }

We show that an optimal probe state among strategies employing the
same average energy is the squeezed vacuum. It was previously shown
\cite{smerzi,furusawa} that an optimal probe state is also a highly
excited Fock state (which is much more complicated to create,
impossible with current technologies). We also show the optimal
detection strategy: an anti-squeezing transformation (i.e.~a squeezing
in the orthogonal direction), followed by a photodetection. Current
experiments and proposals use homodyne detection,
e.g.~\cite{roni,natureaxion}. We show that, surprisingly, homodyne
detection is not only suboptimal, but even useless in the relevant
limit $\var\to 0$: a result that corresponds to Rayleigh's curse
\cite{mankey,mankey1,mankey2,aephr}, which, after proper formulation,
can be seen as a special case of our theorem. A classical state
(coherent or thermal) with $N$ average photons can only attain at most
the same sensitivity of the vacuum $|0\>$, whereas employing a
squeezed or a Fock state we find a quantum Fisher information
proportional to $N$ asymptotically for small $\var$, which proves a
$\sqrt{N}$ enhancement.


We start by providing the precision limits of all spreading channels
and showing a simple strategy that attains those limits: a simple
yes-no projection onto the initial state of the probe.  For the
channels coming from averaging of unitary transformations rotated over
additional parameter, we show that the Fisher information of the
averaged output state is equal to averaged Fisher information
calculated for the pure states, so no information is lost.  Finally,
we study in detail the example of the estimation of a displacement in
a channel where the displacement phase is completely randomized.

\section{Optimality of self-projection measurement}
Consider the family of channels depending on the unknown positive
parameter $\var\geq 0$ satisfying
$\lim_{\var\to 0}\Lambda_{\var}[\rho]=\rho$ (i.e.~they become the identity when
$\var=0$).

The aim is to estimate the exact value of $\var$ by using a probe system in the input state $\rho$ (which may itself be composed of $N$ entangled sub-probes) and performing the measurement $\{\Pi_i\}$ on the output state $\rho_\var=\Lambda_\var[\rho]$. This results in a probability distribution $p(i|\var)=\tr(\Pi_i\rho_\var)$. After $M$ repetitions we assign the estimator to the sequence of the measurement results -- $\tilde\var(i_1,i_2,...,i_M)$. From the Cramer-Rao bound (CR), for any unbiased estimator, the RMSE is bounded from below as
\begin{equation}
\Delta\tilde\var\geq \frac{1}{\sqrt{M}\sqrt{F_C(\rho_\var,\{\Pi_i\})}},
\end{equation}
where $F_C(\rho_\var,\{\Pi_i\})$ is the classical Fisher information (CFI) (which may depend on number of entangled probes $N$ used in each repetition). The CR inequality is known to be asymptotically saturable in the limit $M\to\infty$; in practice the amount of necessary repetitions $M$ depends on the specific model.

For a given output state $\rho_\var$, the maximal value of the  classical Fisher information is equal to the  quantum Fisher information (QFI)
\begin{equation}
\label{clasq}
\max_{\{\Pi_i\}}F_C(\rho_\var,\{\Pi_i\})=F_Q(\rho_\var):=\tr(\rho_\var L^2),
\end{equation}
where $L$ is symmetric logarithm derivative $\frac{d\rho_\var}{d\var}=\frac{1}{2}(L\rho_\var+\rho_\var L)$. 

Since the channel is a linear map and the QFI is a convex function, the optimal state for estimating $\var$ is a pure state $\rho=\ket{\psi}\bra{\psi}$. Below we show, that for any spreading channel, the simple projection measurement on the initial state, i.e.
\begin{equation}
\label{measurement}
   \Pi_0=\ket{\psi}\bra{\psi}, \Pi_1=\openone-\Pi_0
\end{equation}
saturates \eqref{clasq} for small values of $\var$. More precisely, we assume that $\lim_{\var\to 0^+}F_Q(\rho_\var)$ converges to a fixed value $F_Q^+(\rho_0)$ (see App.~\ref{s:dis} and \cite{zhou} for a broader technical discussion, which shows that this assumption is inconsequential as long as the value $\var$ is guaranteed to be non-negative). Then we show that
\begin{equation}
\label{fcc}
   F_C(\rho_\var,\{\Pi_i\})=F_Q^+(\rho_0)+\mathcal O(\var).
\end{equation}
Proof: the  probability of successful projection of the final state $\rho_\var$
onto the initial state, relative to the POVM element $\Pi_0$ is equal to $p_0(\var)=\tr(\rho_\var\ket{\psi}\bra{\psi})$. Using the relation between QFI and the fidelity via the Bures metric,
for small $\var$ we get
\begin{equation}
\label{bures}
\tr(\rho_\var\ket{\psi}\bra{\psi})=1-\frac{1}{4}F_Q^+\var^2+\mathcal O(\var^3),
\end{equation}
so
\begin{equation}
p(0|\var)=1-\frac{1}{4}F_Q^+\var^2+\mathcal O(\var^3),\quad
p(1|\var)=\frac{1}{4}F_Q^+\var^2+\mathcal O(\var^3).
\end{equation}
The CFI for this distribution is
\begin{multline}
   F_C(\rho_\var,\{\Pi_i\})=\sum_{i=0,1}\tfrac{1}{p(i|\var)}(\tfrac{\partial}{\partial\var}p(i|\var))^2=\\
   \tfrac1{p(1|\var)[1-p(1|\var)]}\left(\tfrac\partial{\partial\var}p(1|\var)\right)^2
=F_Q^+(\rho_0)+\mathcal O(\var),
\label{fcc1}
\end{multline}
which ends the proof.

Two major issues connected with the measurement \eqref{measurement}
should be mentioned. Namely, the smaller is $\var$, the larger is the
number of repetitions $M$ needed to saturate the CR bound, and also
the more susceptible to noise is the measurement
\cite{kurdzialek2022measurement}. For example, consider the noise
which changes the probabilities as
$p_0(\var)\to (1-\epsilon)p_0(\var)+\epsilon/2$ and
$p_1(\var)\to (1-\epsilon)p_1(\var)+\epsilon/2$. Then, from
\eqref{fcc1} one can see that in the limit $\var\to 0$ even arbitrary
small $\epsilon$ may completely decrease the value of Fisher
information. In practice, the procedure works well and gives good
estimates of $\var$ if the noise $\epsilon\ll p_1(\var)$ and the
number of repetitions $M\gg 1/p_1(\var)$. Note that this issue
  is not a specific defect of this protocol, but rather an unavoidable
  difficulty: for certain types of models, maximizing QFI is
  unavoidably connected with extreme sensitivity to
  noise~\cite{kurdzialek2022measurement}.

A physically relevant class of spreading channels is the one obtained by averaging some unitary channel over the direction of action. We consider a quantum channel of the form:
\begin{equation}\label{cha}
    \Lambda_\var(\rho)=\int d\varphi\:p(\varphi)\: U_{\var,\varphi}\: \rho\: U_{\var,\varphi}^\dagger,\qquad
    U_{\var,\varphi}=U^\dagger_\varphi e^{i\var G} U_\varphi.
\end{equation}
This means that, every time the channel is used, $\varphi$ is
independently randomly drawn from the distribution $p(\varphi)$. This
is not a randomization of the parameter $\varphi$ of the probe, but of
the channel itself. Assuming a pure input state, from convexity 
\begin{equation}
\label{mean}
    \forall_{\var>0}\,F_Q(\rho_\var)\leq \int d\varphi\:p(\varphi) F_Q(\ket{\psi_\var^\varphi}):=\overline{F_Q}(\var),
\end{equation}
where $\ket{\psi_\var^\varphi}\equiv U_{\var,\varphi}\ket{\psi}$ and
$\overline{F_Q}$ is the average QFI of $|\psi_\var^\varphi\>$ over
$\varphi$. Below we show, that this inequality is tight in the limit of small $\var$ -- indeed, for fixed initial state, no information is lost during averaging over $\varphi$. Moreover, all this information may be extracted through a projection on the initial state. Indeed, from the Bures metric for pure states we have:
\begin{equation}
|\braket{\psi|\psi_\var^\varphi}|^2=1-\frac{1}{4}F_Q(\ket{\psi_\var^\varphi})\var^2+\mathcal O(\var^3),
\end{equation}
which, after averaging over $\varphi$, gives \eqref{bures} with $\overline{F_Q}(\alpha)$ in place of $F_Q^+(\rho_0)$, which shows that they differ by the factor $\overline{F_Q}(\alpha)-F_Q^+(\rho_0)=
\mathcal O(\var)$

Note that, if $p(\varphi)$ is not uniform, some care must be taken to
estimate $\var$ from the measurement outcomes, see App.~\ref{s:est}.

\section{Optical displacement  estimation}
\label{sec:opbound}

\begin{figure}[h!]
\includegraphics[width=8cm]{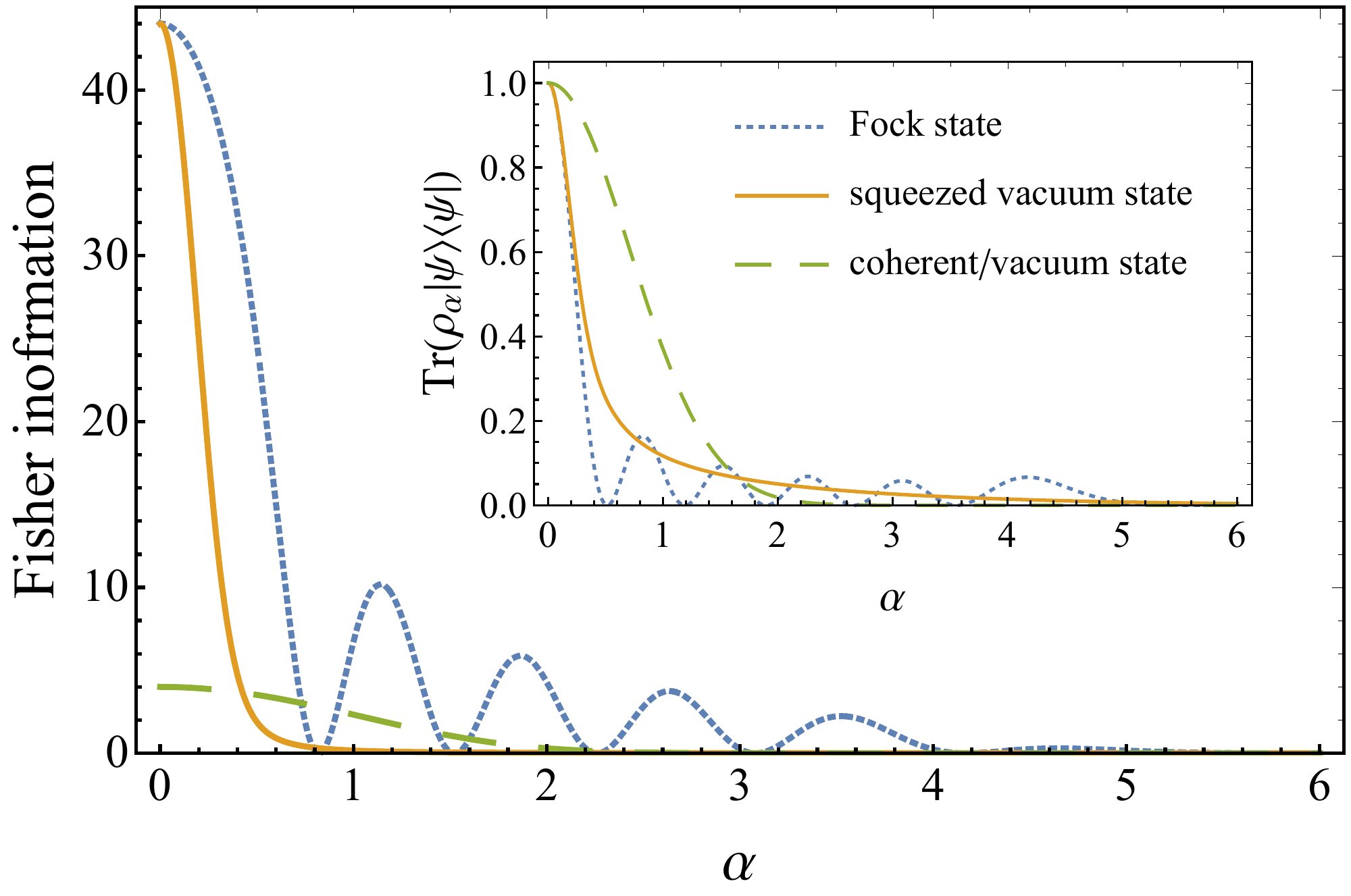}
\caption{Classical Fisher information for the measurement
    \eqref{measurement} and the fidelity (inset) between the initial
  state $|\psi\>$ and the final state
  $|\psi_\var^\varphi\>=D(\var,\varphi)|\psi\>$ (with $D$
  displacement), averaged over $\varphi$ as a function of $\var$. Blue
  dotted - Fock state, yellow solid - squeezed vacuum state, green
  dashed - coherent state, all with $N=5$ average photons. For the
  vacuum state the fidelity is the same as for coherent state (green
  dashed). }\label{f:fid}
\end{figure}

In this section we consider a physically relevant example of spreading
channels: the estimation of the amplitude $\var$ of a displacement
$D(\var,\varphi)= U^\dagger_\varphi e^{i\var G}U_\varphi$, where
$ G=\frac{1}{i}(a^\dagger-a)$ and $U_\varphi=e^{-i\varphi a^\dagger a}$, with
random phase $\varphi$. 
We consider the scenario where the experiment is repeated $M$ times
  with a bound $N$ for the mean input energy in each realization. This
  is related to typical realistic constrains, where the total number
  of repetitions $M$ is restricted by the time of observation, which
  is independent of the amount of resources used in a single
  repetition (i.e.~one cannot reduce $N$ increasing $M$ because of total
  time constrains).
 For simplicity of the notation we introduce $G(\varphi)=U_\varphi^\dagger G U_\varphi=
\frac{1}{i}(e^{i\varphi}a^\dagger-e^{-i\varphi}a)$. For each $\varphi$, the QFI is
\begin{multline}
\label{full}
  F_Q(|\psi_{\var}^\varphi\>)=4(\braket{\psi_{\var}^\varphi|G(\varphi)^2|\psi_{\var}^\varphi}-\braket{\psi_{\var}^\varphi|G(\varphi)|\psi_{\var}^\varphi}^2)=\\
  4(\braket{\psi|G(\varphi)^2|\psi}-\braket{\psi|G(\varphi)|\psi}^2)
    \leq 4\braket{G(\varphi)^2}\\
 =4\braket{-e^{2i\varphi}a^{\dagger2}-e^{-2i\varphi}a^2+2a^\dagger
    a+1},
\end{multline}
where in the second equality we used the fact that acting with $e^{i G(\varphi)}$ does not change the variance of $G(\varphi)$. Therefore, the average is upper bounded by the average photon number of the initial state as
\begin{equation}
\label{bound}
  \overline{F_Q}=\frac{1}{2\pi}\int d\varphi\:F_Q(|\psi\>_\var^\varphi)\leq 8(\braket{a^\dagger a}+\frac{1}{2})
\end{equation}
which was derived in a different manner in \cite{smerzi}.
This bound can be clearly saturated by a Fock state
\cite{smerzi} which is invariant for
$U_\varphi=e^{i\varphi a^\dag a}$. What was unknown up to now is that
it can also be saturated by a squeezed vacuum state $|r,0\>$ ($r$ the
squeezing parameter), where for $\var\sim 0$ we have
\begin{equation}
  F_Q(|\psi_\var^\varphi\>)=8(\cos(2\varphi)\cosh(r)\sinh(r)+\sinh^2(r)+{1}/{2}),
\label{cffi}\end{equation}
where $N=\<a^\dag a\>=\sinh^2(r)$, so indeed after averaging over
$\varphi$,  \eqref{bound} is saturated.  As can be expected,
  for the squeezing in direction of the shift ($\varphi\approx 0$),
  the QFI is significantly enhanced, while for the perpendicular
  direction ($\varphi\approx \pi/2$) it performs even worse than the
  vacuum state. The intuition behind our procedure is that these
  competing effects do not cancel (due to the nonlinearity of the QFI)
  and, after averaging over $\varphi$, the bound \eqref{bound} is 
  saturated.  For the optimal measurement proposed in
  \eqref{measurement} the probability of outcome $0$ is exactly given
  by the fidelity between the initial and final state averaged over
  $\varphi$ (the averaging being irrelevant for Fock states). For the
squeezed state \cite{brask2021gaussian}:
\begin{multline}
\label{avv}
  \int\frac{d\varphi}{2\pi}|\braket{r,0|D(\var,\varphi)|r,0}|^2=\\
  \int\frac{d\varphi}{2\pi}e^{-\var^2(\cos^2(\varphi)e^{2r}+\sin^2(\varphi)e^{-2r})}=
\\
  e^{-\var^2\cosh(2r)}I_0(\var^2\sinh(2r))=\\
  e^{-\var^2(2N+1)}I_0(\var^22\sqrt{N(N+1)}),
\end{multline}
with $I_0$ modified Bessel functions of the first kind, whereas for Fock states \cite{oliveira1990properies,smerzi}:
\begin{equation}
  |\braket{N|D(\var,\varphi)|N}|^2=e^{-|\var|^2}(\mathcal L_N(|\var|^2))^2,\ \forall\varphi
\label{acq}
\end{equation}
with $\mathcal L_N$ the Laguerre polynomial. Both behave in the same
way for small $\var$. So, while both states are equally optimal in the limit $\var\to 0$, for
  larger values of $|\var|$ the FI for the Fock state is typically
  higher even though at specific points it is null (see
  Fig.~\ref{f:fid}) and there Fock states become useless. So, in the
  case of local estimation, the Fock state performs better typically,
  but the situation changes for global estimation since the averaged
  squeezed state has a monotonic decreasing fidelity, the Fock state
  does not. So, the value of $\var$ cannot always be derived uniquely
  solely from the measurement \eqref{measurement} if one uses a Fock
  state, but it can in the case of squeezed states. Indeed, if
one uses a squeezed vacuum, the maximum likelihood estimator $\tilde\var_{\t{ML}}(m_0,m_1)$ (where $m_{0,1}$ are the
number of measurements with outcome $0,1$ in the measurement
\eqref{measurement}) is simply
given by the inverse of the function
$p(0|\var)$ (with respect to $\var$) at point $p(0|\var)=\frac{m_0}{m_0+m_1}$. Clearly, when a finite number of measurements are
employed, statistical fluctuations in the average \eqref{avv} will
become important, which can be estimated through Monte-Carlo methods
(App.~\ref{s:mc}). Up to now, we considered the case where the phase
is completely randomized between different measurements. If there is
no randomization and the value of $\varphi$ is known, then the optimal
strategy is known \cite{optsq}: use a squeezed vacuum with
$\varphi$-dependent squeezing. The intermediate case in which the
randomization happens slowly is analyzed in App.~\ref{s:ran}.

The quantum enhancement can be shown if one starts from a coherent
state $|\beta\>$ with average number of photons $|\beta|^2$, it is
clear that
\begin{align}
|\<\beta|D(\var,\varphi)|\beta\>|^2=|\<0|D(\var,\varphi)|0\>|^2=e^{-\var^2}
\labell{coh}\;,
\end{align}
namely, the fidelity between a coherent state and a displaced
  coherent state is the same as the fidelity between the vacuum and
  the displaced vacuum for the same degree of displacement.  So one
can obtain a strategy that performs equally well using a zero-energy
vacuum state instead of a coherent state (and both are impervious to
the value of $\varphi$ which does not affect the fidelity). In
  the context of \eqref{bound}, the inequality is not saturated in
  this case, since for general coherent state $\ket{\beta}$ the term
  $\braket{\beta|G(\varphi)|\beta}$ (appearing in \eqref{full}) has
  non zero value.

Thermal states will perform worse. So, it is clear that any
classical strategy of energy $|\beta|^2=N$ is, at best, as effective
as a strategy that uses a vacuum state (which is the optimal strategy
for zero energy). In contrast, \eqref{bound} shows that the optimal
quantum strategy of energy $N$ has QFI of order $N$. While this
resembles the typical $\sqrt{N}$ enhancement in precision of quantum
strategies vs. classical ones in quantum metrology, interestingly it
has a different origin than usual: it originates from the fact that
any energy devoted to classical strategies is completely useless,
rather than from a different allocation of the resources.

Both Fock state and squeezed states may help with the problem with
  noise susceptibility discussed after  \eqref{fcc1}. Indeed, looking at \eqref{avv} 
  one can see that for large N the probability of getting result 0 is
  approximately a function of $\sim\alpha^2N$ , so even for
  extremely small values of the parameter one is able to keep the
  probability of the result 1 sufficiently large, by increasing the energy.

We now comment on the practicalities of the two strategies. On one
hand, the Fock state strategy
\cite{oliveira1990properies,smerzi} requires that the
initial state of the radiation be prepared in $|N\>$. There is no
currently known technique to prepare such state for the
electromagnetic field which is scalable to high values of $N$.  Then,
at the detection stage one must project onto the initial state $|N\>$,
which can be implemented with a photon-number resolving photodetector.
Such devices exist, but cannot retain effective photon-number
resolution to large numbers $N$ of photons. On the other hand, the
squeezed state strategy seems more practical, since there is a vast
literature for the preparation of squeezed states at different
wavelengths spanning from the optical \cite{opticalsq} to the
microwave \cite{microwsq}. At the detection stage, one must evaluate
the probability that the output state of the channel
$|\psi_\var^\varphi\>$ is equal to a squeezed vacuum, namely
$p(0|\var)=|\<0|S^\dag(r)|\psi_\var^\varphi\>|^2$, where
$|r,0\>=S(r)|0\>$. This expression can be also interpreted as the
calculation of the overlap between the state
$S^\dag|\psi_\var^\varphi\>$ with the vacuum state $|0\>$, which can
be easily implemented: the first is the state obtained by applying the
inverse squeezing transformation $S^\dag$ {\it after} the channel and
then performing a photodetection. The probability to obtain zero
photons at the measurement will give $p(0|\var)$, whereas the
probability to obtain one or more photons will give $p(1|\var)$.  An
avalanche photodiode (APD) or any equivalent avalanche photodetector
(e.g.~transition edge sensors) will provide such output signal in the
ideal case.  The whole procedure must be gated so that it is clear
when a ``no click'' must be interpreted as an outcome.

It might seem surprising that one must un-squeeze the signal before
the detection, since the state preparation involves a squeezing
transformation: however, the squeezing, channel application and
un-squeezing is equivalent to a (sub-shot noise) effective
amplification of the quadrature, whenever the signal is orthogonal to
the squeezing. In general, this condition is not warranted but, as
shown in the previous section, this is irrelevant: after the averaging
over $\varphi$, this procedure still performs very well, and it
performs optimally for small values of $\var$, which is the regime
of interest.

One could think that an alternative detection for the squeezed
strategy could be implemented through homodyne detection, by measuring
the quadrature of the light consistent with the squeezing phase. While
this strategy may be useful in some cases, surprisingly in the regime
of small $\var$, this strategy fails in a way reminiscent of
Rayleigh's curse \cite{mankey,aephr} for the evaluation of the
distance of two point sources (while the last problem also may easily described within our formalism by taking $\ket{\psi}\propto\exp(-x^2/4\sigma^2)\ket{x}$, $G=\frac{1}{i}\partial_x$, $p(\varphi)=\tfrac{1}{2}(\delta(\varphi)+\delta(\varphi-\pi))$; see
App.~\ref{s:rayl} for broader discussion about relation between these two models).

\section{Conclusions}
We have provided the optimal estimation strategy for the spreading
parameter $\var$ of spreading noisy channels, where some other
parameter $\varphi$ is randomized. This is one of the very few
instances where we can give the optimal metrology strategy for a noisy
channel. We analyzed a specific instance of spreading channels: the
estimation of the amplitude of a displacement with random phase. We
show a quantum enhancement equal to the square root of the number of
photons employed in the estimation and we derived a new optimal
strategy, based on squeezed vacuum states, that is practically
implementable with current techniques, in contrast to the previously
known one \cite{smerzi,furusawa} based on Fock states.

This material is based upon work supported by the U.S. Department of
Energy, Office of Science, National Quantum Information Science
Research Centers, Superconducting Quantum Materials and Systems Center
(SQMS) under contract number DE-AC02-07CH11359. W. G. acknowledges
support from the National Science Center (Poland) Grant No.
2020/37/B/ST2/02134 and the Foundation for Polish Science (FNP) via
the START scholarship. We acknowledge useful feedback from M. Sacchi
and R. Demkowicz-Dobrza\'nski.


\appendix

\section{Discontinuity of QFI at $\var=0$}
\label{s:dis}
In this appendix we recall a simple example from \cite{zhou} to show
that the issues mentioned in \cite{zhou} do not affect our reasoning
in the case we are interested in. Namely, the situation when the
parameter $\var$ to be estimated is non-negative.

Consider the family of states:
\begin{equation}
    \rho_\var=\var^2\ket{0}\bra{0}+(1-\var^2)\ket{1}\bra{1}
\end{equation}
and let $\var$  be an arbitrary real number. Then for any point different from $\var=0$, the  symmetric logarithmic derivative is equal to
\begin{equation}
   L=\tfrac{2}{\var}\ket{0}\bra{0}-\tfrac{2\var}{1-\var^2}\ket{1}\bra{1}),
\end{equation}
which leads to a QFI
\begin{equation}
    F_Q(\rho_\var)=\tr(\rho_\var L^2)=4+\frac{4\var^2}{1-\var^2},
\end{equation}
so $\lim_{\var\to 0^+}F_Q(\rho_\var)=4$. However, for $\var=0$, the
symmetric logarithmic derivative is simply $L=0$, so also
$F_Q(\rho_\var)|_{\var=0}=0$. The simple interpretation of this fact is
that the state $\rho_\var$ does not distinguish between positive and
negative values of $\var$, so any measurement performed on it cannot
allow for local estimation of $\var$ around point $\var=0$ (while for
any different point $\var\neq 0$ local estimation will be possible).
It is clear that, in our case, where we consider only non-negative
$\var$, this issue does not appear, so it is reasonable to consider
$F^+_Q(\rho_0)$ instead of $F_Q(\rho_\var)|_{\var=0}=0$.

\section{Estimator of $\var$}
\labell{s:est}
In order to recover the value of $\var$, one must know the actual form
of the probability distribution $p(\varphi)$, since the maximum
likelihood estimator is obtained from the theoretical probability of
obtaining the results which, in turn, depends on $p(\varphi)$.
Indeed, the maximum likelihood estimator is \begin{equation}
   \tilde\var_{ML}(k_0,k_1):=\t{arg max}_{\var}p(k_0,k_1|\var),
\end{equation}
where $k_{0/1}$ the number of results $0/1$. Therefore, if one has no
a priori knowledge on $p(\varphi)$ we cannot simply assume that the
distribution is uniform, as it may lead to a strongly biased
estimator.  However, one can add an additional random-uniform
rotation, so that the overall rotation becomes uniformly distributed.
Indeed, assume that $\varphi$ is non-uniform distributed with some
unknown function $p(\varphi)$. Add an additional rotation of the
initial state $U_\theta\ket{\psi}$ as well as a final rotation prior
to the measurement $U_\theta \Pi_0 U_\theta^\dagger$. Then,
\begin{multline}
\label{randomize}
  p(0|\var)=\int p(\varphi)d\varphi \int_0^{2\pi}\frac{d\theta}{2\pi}|\braket{\psi|\psi_\var^{\theta+\varphi}}|^2=\\
  \int_0^{2\pi}\frac{d\theta}{2\pi}|\braket{\psi|\psi_\var^{\theta}}|^2,
\end{multline}
which leads to the same statistics as an initial uniform distribution
of $\varphi$. Of course, this is in general a suboptimal strategy, as
the mean value of the QFI for uniform distribution may be smaller than
the one for the true $p(\varphi)$. In practice, it will usually be
better to spend some resources to find an estimate to $p(\varphi)$ and
then use the rest to estimate $\var$.

\section{Performance of different types of initial states}
\label{s:mc}

In this section we analyze the performance of other initial states, in
addition to the squeezed vacuum and Fock states considered in the main
text. Moreover, we present some Monte-Carlo simulations that show how
the squeezed vacuum (and the other states) perform when considering a
finite number $M$ of repetitions of the experiment with randomly
varying phase $\varphi$ (in the main text we considered the simple
case in which the randomization is perfect).

\begin{figure}[ht]
 \epsfxsize=1.\hsize\leavevmode\epsffile{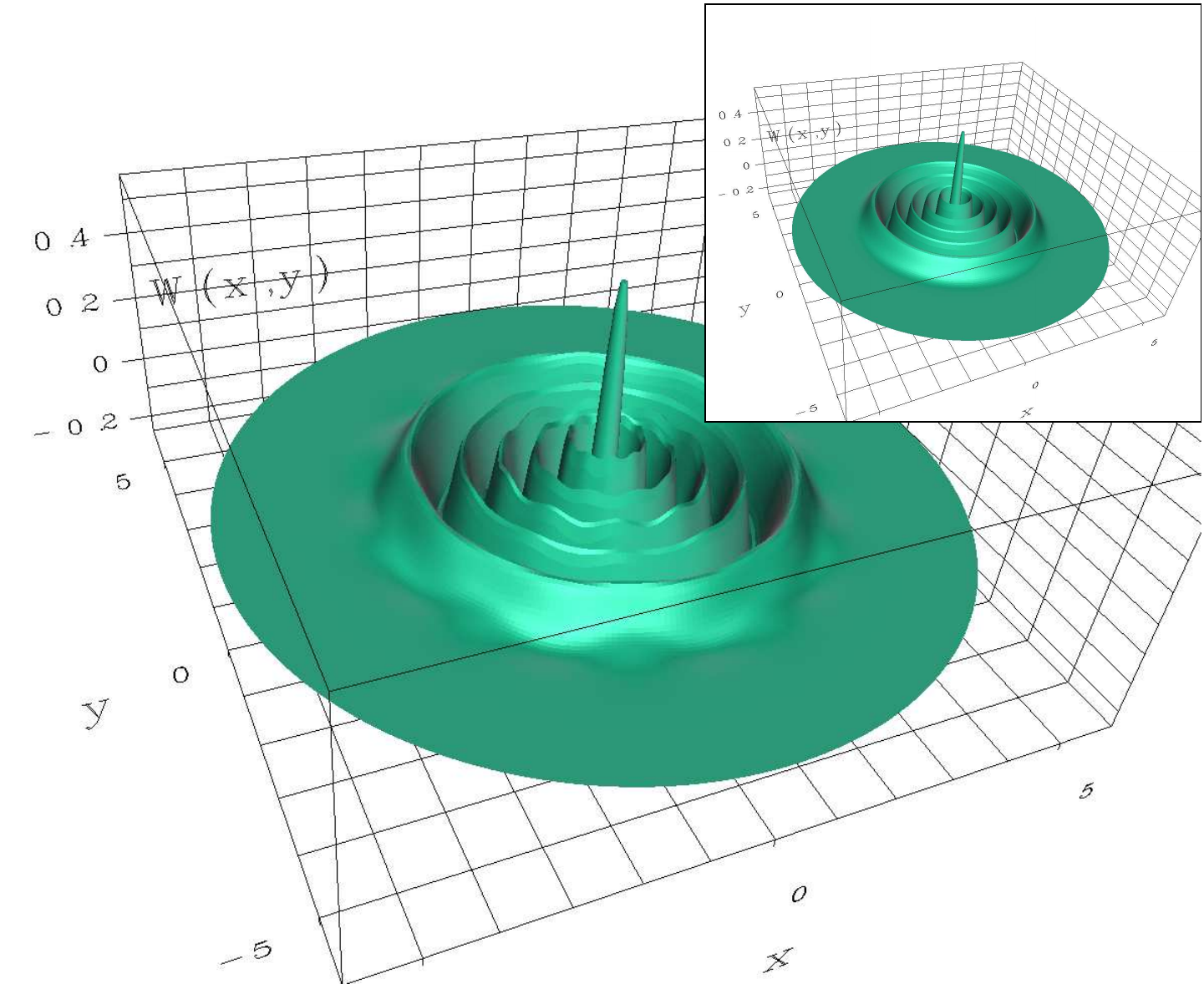}
 \caption{Wigner function of the multi-cat $  |\mbox{cat}_{N,\theta}\>$ with $N=10$ and 10 average
   photons, i.e.~$\beta\simeq 8.3045$ in Eq.~\eqref{multicat}. It is
   evident that such a state closely approximates a Fock state with
   the same number of photons (inset).
   \label{f:wignercat}}\end{figure}

Consider the multi-cat state \cite{zurek}
\begin{align}
  |\mbox{cat}_{N,\theta}\>\propto \sum_{n=1}^N|\exp(2i\pi n/N+i\theta)\beta\>
\labell{multicat}\;,
\end{align}
namely a superposition of $N$ coherent states with equispaced phases.
As shown in Fig.~\ref{f:wignercat}, this is a good approximation for
the Fock state already for moderate values of $N$, and it may be
simpler to create in practice: at least for $N=2$ it can be created in
quantum optics, e.g.~\cite{nostro,kittens} or in the microwave regime
also for $N=3,4$ \cite{science}.

\begin{figure}[ht]
 \epsfxsize=1.\hsize\leavevmode\epsffile{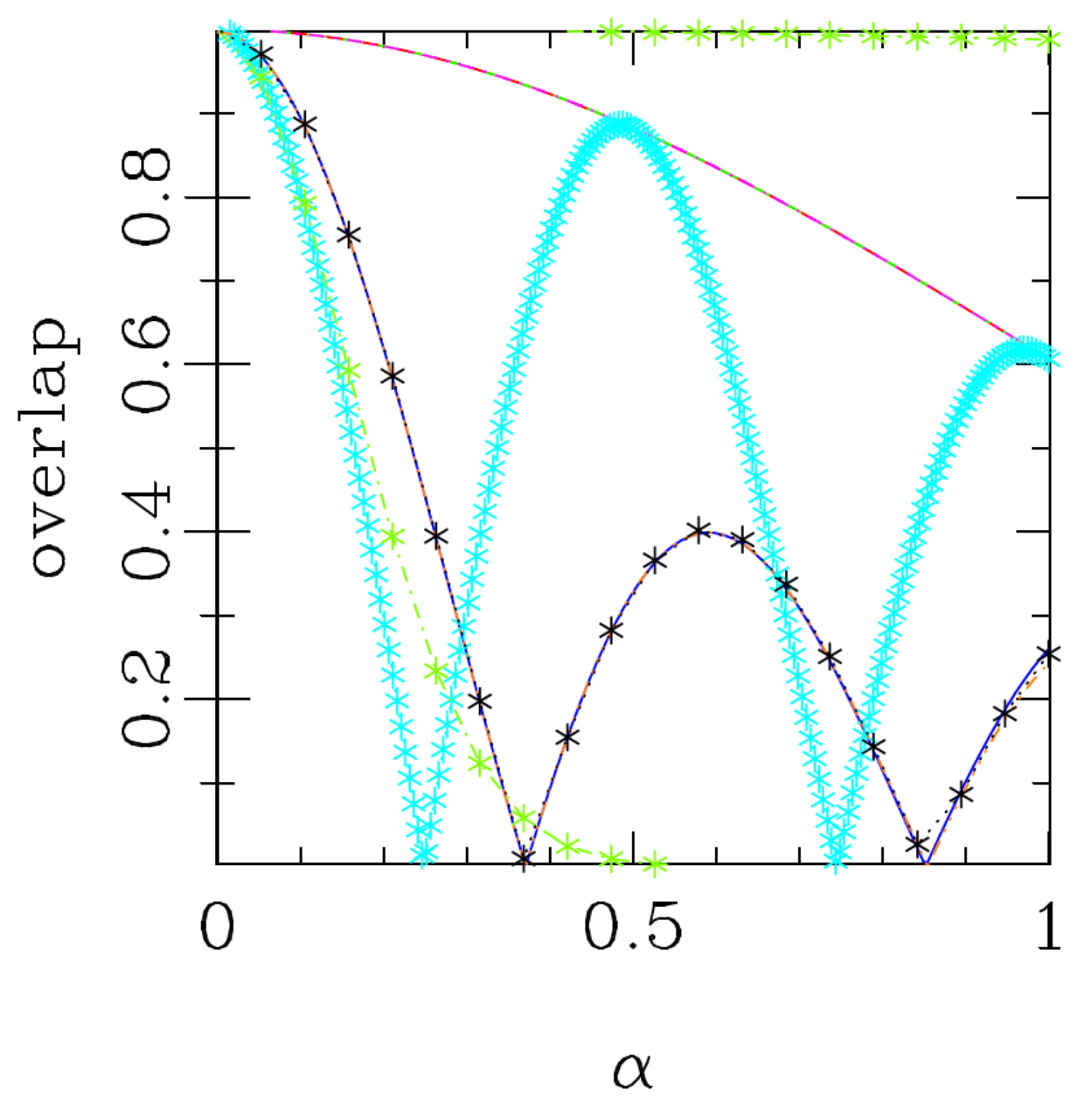}
 \caption{Plot of the overlap between the initial state and the
   initial state displaced by $D(\var,\varphi)$ (here with fixed phase
   $\varphi$) as a function of $\var$. Namely,
   $overlap=|\<\psi|{\cal D}(\alpha)|\psi\>|$ where $|\psi\>$ is any
   of the states we consider here. The squeezed vacuum state
   $|r,\theta\>$ (green stars with dash-dotted line) performs worse if
   the squeezing phase $\theta$ is parallel to the displacement phase
   $\varphi$ (upper curve) and it performs very well if the two phases
   are orthogonal (lower curve). The coherent state, the vacuum state
   and the two-cat state with phase $\theta$ parallel to the
   displacement all perform equally (higher solid multi-colored
   curve). They all perform quite badly: the overlap decreases very
   little as a function of $\alpha$. The multi-cat state $  |${cat}${}_{N,\theta}\>$ (here we use
   $N=10$ cats) is plotted as the continuous blue curve for parallel
   phase $\theta$ and as the orange dot-dashed curve for orthogonal
   phase $\theta$ (they are indistinguishable in this plot as there is
   little difference between their performance: the multi-cat is
   almost insensitive to phase).  In contrast, the two-cat state is
   quite sensitive to the phase $\theta$: the cyan dotted curve plots
   its performance when its phase is orthogonal to the displacement
   (good behavior), whereas it behaves exactly as the coherent state
   if the phase is parallel. The best state overall appears to be the
   2-cat state (but is highly phase-sensitive). In contrast, the Fock
   state (black line) is basically identical to the multi-cat and are
   both independent on the phase (the Fock is exactly independent, the
   multi-cat is approximately independent).  All states here (except
   for the vacuum state) have the same average number of photons,
   $\<a^\dag a\>=10$.
   \label{f:fig}}\end{figure}

\begin{figure}[ht]
 \epsfxsize=1.\hsize\leavevmode\epsffile{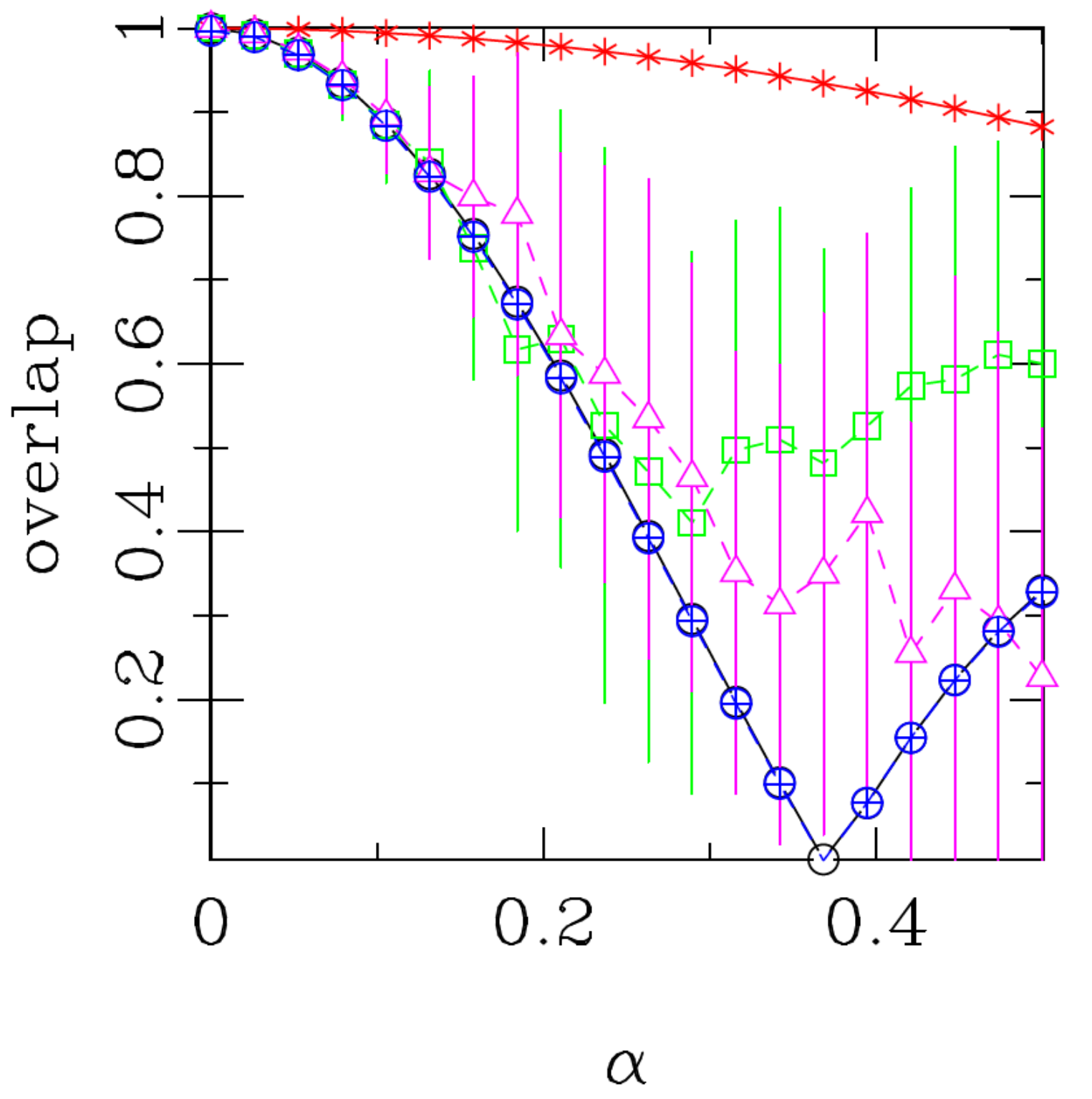}
 \caption{Monte-Carlo simulation of the overlap between the initial
   state and the initial state displaced by $D(\var,\varphi)$ with
   randomly chosen $\varphi$ as a function of $\var$. The coherent
   state/vacuum (red stars), the Fock state (black lines) and the
   10-cat (blue circles) are unaffected (the multi-cat is affected
   slightly, but the effects cannot be seen at this scale). The
   two-cat (green squares) is quite phase sensitive and the
   fluctuations for random phases can be easily seen (vertical bars).
   The squeezed vacuum (pink triangles) is also very sensitive to
   phase. In the regime $\var\to 0$, the Fock state, the squeezed
   vacuum, the two-cat and the 10-cat all perform in a similar way.
   The vertical bars are not the statistical error bars: they are an
   estimation of the fluctuations (the root mean square of the
   overlap). [The error bars would be the root mean square divided by
   the square root of the number of simulated phases.] Here all states
   (except the vacuum) have $\<a^\dag a\>=10$ and an average over
   $M=50$ uniformly distributed random values of $\varphi$ is
   performed.
   \label{f:figmc}}\end{figure}

In Fig.~\ref{f:fig} we consider various states: the coherent state
$|\beta\>$, the vacuum state $|0\>$, the squeezed state $|r,\theta\>$,
the multicat $ |\mbox{cat}_{N,\theta}\>$ for different values of $N$
and $\theta$. We plot the overlap between each of these states and its
displaced version after a displacement $D(\var,\varphi)$ with a fixed
phase. A good metrological state is the one that is highly sensitive
to $\var$, namely whose value changes rapidly as a function of $\var$.
In Fig.~\ref{f:figmc} we present a Monte-Carlo simulation that shows
how the randomly varying $\varphi$ affects the overlap. As expected
(see also Fig.~\ref{f:fid}), the random phase ruins the metrological
sensitivity of the more phase-sensitive states. Yet, in the regime of
small $\var$ we are interested in, the squeezed state $|r,0\>$ and the
two-cat state $ |\mbox{cat}_{N=2,\theta}\>$ still behave as well as
the Fock state which is insensitive to $\varphi$.

\section{Slowly varying $\varphi$}
\label{s:ran}

Up to now we have assumed that the rotation $\varphi$ changes randomly
from one shot to the next. In practice, there might be a slow time
dependence such that the value of $\varphi$ is reasonably constant for
some time: consider a time-dependent probability $p(\varphi|{t})$,
which gives $p(\varphi)$ of \eqref{cha} when averaged over a large
time interval, but which is sufficiently tight for a short period of
time. Then one can use a feedback strategy in which one quickly
performs an estimate of $\varphi$ and then uses it to prepare a
squeezed vacuum with squeezing parameter orthogonal to $\varphi$. Such
a state performs better than the Fock state, as long as the value of
$\varphi$ is known. Alternatively, instead of first estimating
$\varphi$ and then using its value in the subsequent measurements, one
could also employ multiparameter estimation \cite{matteo}.  

In essence, the bounds to precision presented in this paper can be
beaten under the hypothesis that the random parameter $\varphi$
changes sufficiently slowly.

\togli{ it may be sometimes reasonable
to assume that this probability changes with time $p(t,\varphi)$,
where for fixed $t$ is has rather tight shape, while after integration
over long time $t$ it becomes uniform.
Then, for any fixed time $t$:
\begin{equation}
    \braket{\hat x(t)}=\int p(t,\varphi)d\varphi\cos(\varphi)\sqrt{2}\var\neq 0.
\end{equation}
Therefore, if $p(t,\varphi)$ changes slowly with time (e.i. it is possible to perform many repetition of the experiment unless it change significantly), even if the form of $p(t,\varphi)$, one may perform regularly the measurements of both $\hat x$ and $\hat p$, keeping in memory the time when each measurement was performed
(ideally, one would like to prepare different states for measuring $\hat x$ and $\hat p$, squeezed in mutually perpendicular direction). Then from $\braket{\hat x(t)}$ and $\braket{\hat p(t)}$, they can estimate current phase $\varphi(t)$, which allow for estimating the value of $\var$ even in the limit $\var\to 0$.}

\section{Quadrature measurement and Rayleigh's curse}
\label{s:rayl}
In this section we analyze how a quadrature measurement compares to
the projection onto the initial state that was considered in the main
text. Surprisingly, we show that, even though this measurement
strategy performs well for large values of $\var$, it does not work
well in the regime of asymptotically small $\var$ we are interested
in. 

This failure mode is somewhat reminiscent of the Rayleigh's curse in
discriminating two point-like sources at a distance using a
finite-aperture lens. It is known that using a quantum optimized
detection strategy, the Rayleigh's curse can be beaten \cite{mankey}.
Analogously, in the estimation of $\var$ in the regime $\var\to 0$,
the quadrature measurement fails, whereas the projection onto the
initial state is optimal. 

Consider estimating the value of $\var$ from measuring the quadrature
of the light in a direction consistent with the initial squeezing.
I.e.~for an initial squeezed state with zero phase, consider the
measurement of the zero-phase quadrature, namely
\begin{equation}
    \hat x=\frac{1}{\sqrt{2}}(a+a^\dagger)
\end{equation}
Since $\varphi$ is uniformly distributed, the expected value of the
quadrature is $\braket{\hat x}=0$, so $\var$ cannot be simply
estimated from its mean value. Can we use the second moment or the
variance $\braket{\hat x^2}$?  Consider the observable $A=\hat x^2$.
On the squeezed vacuum, displaced and averaged over $\varphi$, we find
\begin{equation}
   \braket{A}=\var^2+\frac{1}{2}e^{-2r},\ 
   \braket{\Delta^2A}=e^{-2r}\left(2\var^2+\tfrac{1}{2}e^{-2r}\right),
\end{equation}
Error propagation gives:
\begin{equation}
   \Delta^2\tilde\var=\frac{\braket{\Delta^2A}}{|\partial_\var \braket{A}|^2}=\frac{e^{-2r}(4\var^2+e^{-2r})}{8\var^2}.
\end{equation}
For $\var^2\gg e^{-2r}$ we find
$\Delta^2\tilde\var\approx \frac{1}{2}e^{-2r}$. It means that the variance scales as the inverse of the mean energy of the input state, which is the  optimal  scaling, as shown in \eqref{bound}. However, for $\var^2\ll e^{-2r}$, the variance scales
inversely to $\var^2$ and goes to infinity for $\var\to 0$. This is
closely related to the Rayleigh curse \cite{mankey}. In fact, in our
problem the output state may be written in following form:
\begin{multline}
    \rho_\var=\int_0^{2\pi}\frac{d\varphi}{2\pi}\ket{\psi_\var^\varphi}\bra{\psi_\var^\varphi}=\\
    \int_0^{\pi}\frac{d\varphi}{\pi}\frac{1}{2}\left(\ket{\psi_\var^\varphi}\bra{\psi_\var^\varphi}+\ket{\psi_\var^{\varphi+\pi}}\bra{\psi_\var^{\varphi+\pi}}\right)=\\
    \int_0^{\pi}\frac{d\varphi}{\pi}\frac{1}{2}\left(\ket{\psi_\var^\varphi}\bra{\psi_\var^\varphi}+\ket{\psi_{-\var}^{\varphi}}\bra{\psi_{-\var}^{\varphi}}\right).
\end{multline}
The convexity of classical Fisher information implies
\begin{multline}
   F_C(\rho_\var,\Pi_x)\leq \\ \int_0^{\pi}\frac{d\varphi}{\pi}F_C\left(\tfrac{1}{2}\left(\ket{\psi_\var^\varphi}\bra{\psi_\var^\varphi}+\ket{\psi_{-\var}^{\varphi}}\bra{\psi_{-\var}^{\varphi}}\right),\Pi_x\right).
\end{multline}
Moreover, for any $\varphi$,
\begin{equation}
\lim_{\var\to 0}F_C\left(\tfrac{1}{2}\left(\ket{\psi_\var^\varphi}\bra{\psi_\var^\varphi}+\ket{\psi_{-\var}^{\varphi}}\bra{\psi_{-\var}^{\varphi}}\right),\Pi_x\right)=0.
\end{equation}

The last equation may be proved for any reasonable initial state (the precise criteria will be given later), not only for squeezed vacuum state. Indeed, for the state $\tfrac{1}{2}\left(\ket{\psi_\var^\varphi}\bra{\psi_\var^\varphi}+\ket{\psi_{-\var}^{\varphi}}\bra{\psi_{-\var}^{\varphi}}\right)$ the probability of getting the result $x$ has the form
\begin{multline}
    p(x|\var)=\tr(\Pi_x\tfrac{1}{2}\left(\ket{\psi_\var^\varphi}\bra{\psi_\var^\varphi}+\ket{\psi_{-\var}^{\varphi}}\bra{\psi_{-\var}^{\varphi}}\right))=\\ \frac{1}{2}(f(x+\sqrt{2}\var\cos(\varphi))+f(x-\sqrt{2}\var\cos(\varphi))),
\end{multline}
where $f(x)=\tr(\Pi_x\ket{\psi}\bra{\psi})$. This implies
\begin{equation}
\label{problim}
 \lim_{\var\to 0}\frac{\partial p(x|\var)}{\partial \var}=0\     \forall x.
\end{equation}
Since $F_C$ is given by:
\begin{equation}
    F_C=\int dx \frac{1}{p(x|\var)}\left(\frac{\partial p(x|\var)}{\partial \var}\right)^2,
\end{equation}
then equation \eqref{problim} does not immediately lead to the conclusion that the CFI vanishes, since we need to pay special attention to the points where also $\lim_{\var\to 0}p(x|\var)=0$ (since in principle it may lead to non-zero Fisher information, as in the example discussed in App. \ref{s:dis}).

Let us restrict to the functions $f(x)$ which have a continuous second derivative with respect to $x$. Moreover, let as assume that there exists an integrable function $g(x)$ such that $\forall_{x,\var}\frac{1}{p(x|\var)}\left(\frac{\partial p(x|\var)}{\partial \var}\right)^2\leq g(x)$ (which is satisfied for both the squeezed vacuum state and the Fock state discussed in this paper). Then, from the dominated convergence theorem \begin{equation}
\label{intlim}
    \lim_{\var\to 0}F_C=\int dx \lim_{\var\to 0}\frac{1}{p(x|\var)}\left(\frac{\partial p(x|\var)}{\partial \var}\right)^2.
\end{equation}
For any point where $f(x)\neq 0$, obviously $\frac{1}{p(x|\var)}\left(\frac{\partial p(x|\var)}{\partial \var}\right)^2=0$. Where $f(x)=0$:
\begin{equation}
\begin{split}
    p(x|\var)&=\cos^2(\varphi)\var^2 \partial^2_x  f(x) +\mathcal O(\var^3)\\
    \partial_\var p(x|\var)&=2\cos^2(\varphi)\var \partial^2_x  f(x) +\mathcal O(\var^2)
\end{split}
\end{equation}
and therefore:
\begin{multline}
    \lim_{\var\to 0}\frac{1}{p(x|\var)}\left(\frac{\partial p(x|\var)}{\partial \var}\right)^2=\\
     \begin{cases}
  0  &\t{if}\, f(x)=0 \\
  4\cos^2(\varphi)\partial^2_xf(x) &\t{if}\, f(x)\neq 0
\end{cases}
\end{multline}
Note, that for any function $f(x)$, which have a continuous second derivative, there could be at least countably many points where simultaneously $f(x)=0$ and $\partial^2_xf(x)\neq 0$, so the value of the integral \eqref{intlim} is equal to $0$ anyway.

In conclusion, in the limit $\var\to 0$, the measurement
\eqref{measurement} is much more informative than measuring
quadrature, which is, in general a poor measurement in this regime.

\end{document}